# Influence of squirt flow on fundamental guided waves propagation in borehole embedded in saturated porous media

Zhi-Wen.CUI*, Jin-Xia.LIU, Ke-Xie.WANG

*Department of Acoustics & Microwaves, Physics Institute, Jilin University,*
*Jie-Fang Road119, Changchun, 130023,China*

Department of Acoustics & Microwaves, Physical Institute, Jilin University, JieFang Road No.119, 130023, Changchun City, P. R.China
*Corresponding author.
E-mail address: *cui.zhiwen@mail.china.com; ljxjlu@sina.com
Tel: 86-0431-8499127(o); fax: 86-0431-8941554

**Abstract**

In this paper, the reservoir is modeled by homogeneous two-phase media based on BISQ model. We focus on the effects of the squirt flow on the fundamental guided waves propagation in borehole embedded in saturated porous media excited by monopole, dipole and quadrupole point sources. The full waveforms acoustic logging in a fluid-filled borehole are simulated. The curves of velocity dispersion, attenuation coefficients and excitation of the fundamental guided waves have shown that velocity dispersions are almost independent of the characteristic squirt flow length, attenuations of guided waves are enhanced due to the squirt flow, and excitations of guided waves are decreased due to the squirt flow. It is possible to estimate the characteristic squirt flow length by attenuation coefficients of the guided waves from acoustical logging data.

***PACS*** 43.20.Jr, 43.20.Mv, 91.60

***Keywords*** Squirt flow; two-phase media; guided waves; borehole; dispersion; attenuation

## 1. Introduction

Both velocity and attenuation have been used to estimate the reservoir parameters such as porosity and permeability from elastic waves in porous rocks. The development of rock acoustics models is important for the interpretation of field data. Biot's theory (1956a,b)[1] has been successfully used to describe some basic behavior of acoustic waves propagating in porous media, but it does not adequately describe the attenuation and dispersion of elastic waves in some reservoir rocks. Dvorkin et.al (1993,1994)[2,3] noticed the "squirt" flow (Mavko and Nur, 1975) [4] which was neglected in Biot's model and incorporated the "squirt" flow mechanism into Biot's theory. The current combined squirt-Biot model is well known as BISQ model. Many scientists have investigated this model, and they extended it to anisotropic rocks [5,6](Parra, 1997, Chen and Yang, 2002). However, only to study the waves in infinite porous media, we cannot know the effects of squirt flow on the guided waves associated with the interface in multiplayer media. The main purpose of this paper is to study the effects of the squirt flow on the guided waves associated with the interface in multilayer media. Actually, the real reservoir usually is assumed to horizontal multilayer or cylindrical multilayer media. We applied this model to simulation of acoustical logging in borehole embedded in saturated porous media (cylindrical multiplayer system). We focus on investigation of the effects of the characteristic squirt flow on the fundamental guided waves propagation in the borehole. Full waveforms computed using the discrete wavenumber method are displayed in the time domain. And the dispersion, attenuation and excitation of the guided waves-the Stoneley waves and the pseudo-Rayleigh mode, the flexural mode and screw mode are analyzed.

## 2. Theoretical formulations

Let ( $r,\theta,z$ ) be the cylindrical coordinate system. The model consists of a fluid filled borehole of radius a, extending to infinity in the z direction, embedded in a radially semiinfinite isotropic porous formation saturated by a Newtonian viscous fluid. We consider multipole point sources defined by Schmitt [7]. The equations of motion in the porous formation are [1,8]

$$\mu_b \nabla^2 \mathbf{u} + (H - \mu_b)\nabla\nabla\cdot\mathbf{u} + C\nabla\cdot\mathbf{w} = \frac{\partial^2}{\partial t^2}\left(\rho\mathbf{u} + \rho_f \mathbf{w}\right) \quad (1)$$

$$\nabla\left(C\nabla\cdot\mathbf{u} + M\nabla\cdot\mathbf{w}\right) = \frac{\partial^2}{\partial t^2}\left(\rho_f \mathbf{u} + \tilde{\rho}\mathbf{w}\right) \quad (2)$$

here $H, M, C, \mu_b$ are the Biot's elastic coefficients that replace the usual Lamé coefficients, their expressions as functions of the bulk moduli of the constitutive grains $K_s$, the skeleton $K_b$, and the fluid $K_f$, the shear modulus of the skeleton $\mu_b$. $\mathbf{w} = \phi(\mathbf{u}_f - \mathbf{u})$ represents the flow of the fluid relative to the solid and measured in terms of volume per unit area of the bulk medium, where $\mathbf{u}_f$ and $\mathbf{u}$ are the absolute displacements of the pore fluid and solid phase of a porous media, respectively. The density of bulk can be expressed by the density of grain $\rho_s$ and the density of pore fluid $\rho_f$ as $\rho = (1-\phi)\rho_s + \phi\rho_f$. When we consider the squirt flow [2], introducing the squirt flow factor $S(\omega, R) = 1 - 2J_1(\lambda R)/\lambda R J_0(\lambda R)$, we can get BISQ model just through replacing $M$ by dynamic modulus $\tilde{M} = MS(\omega, R)$. And there $\lambda = \omega\sqrt{\tilde{\rho}/M}$, $\tilde{\rho} = i\eta/\omega\kappa_0 + \alpha_\infty \rho_f / \phi$, $\eta$ is dynamic viscosity, $\kappa_0$ and $\alpha_\infty$ is the permeability and the tortuosity of porous formation, respectively. $R$ is the characteristic squirt flow length, and is assumed a fundamental property of a rock that can be determined by the experiment [3] (Doverkin et.al. 1994).

Usually, the receivers are listed in the axis of the borehole ($r = 0$). Following Zhang Bi-xing et al (1994)[9](or Schmitt et.al [10]), the generalized pressure can be computed by

$$
\begin{aligned}
P_n(z,t) = {} & \frac{1}{2\pi}\int_{-\infty}^{\infty} V_0(\omega) D_n(\omega) e^{-i\omega t} + \\
& \frac{1}{4\pi^2 n!}\left(\frac{r_0}{4}\right)^n \int_{-\infty}^{\infty}\int_{-\infty}^{\infty} \nu^{2n} A_n(k,\omega) V_0(\omega) e^{i(kz-\omega t)} dk d\omega
\end{aligned}
\qquad (3)
$$

where $V_0(\omega)$ denotes the source spectrum, and $D_n(\omega)$ is the source contribution same as in reference [7]. $A_n(k,\omega) = N_n(k,\omega)/D_n(k,\omega)$ is the generalized reflection coefficient determined by the boundary condition[9]. Here, we only consider the open pore boundary condition. The boundary conditions can be expressed in the matrix form of $D_n(k,\omega)A = 0$, where the elements of the secular matrix $D_n(k,\omega)$ are listed in reference [9] as long as $M$ be replaced by dynamic modulus $\tilde{M}$. The period equation is given as $|D_n(k,\omega)| = 0$. For a certain frequency, it gives the axis-wavenumber of guided wave propagation along the direction of the axis of the borehole. Its root is corresponding to the pole $k_l$ ($l = 1,2,\cdots$) of the reflection coefficient $A_n$. The amplitude of the guided wave excited by a source is given by

$$a_n(k_l,\omega)=\frac{1}{n!}\left(\frac{\nu^2 r_0}{4}\right)^n e^{ikz}\frac{\partial[N_n(k,\omega)/D_n(k,\omega)]}{\partial k}\bigg|_{k=k_l} \quad (4)$$

Any pole corresponds at a given frequency to a complex wavenumber root $[k_l = \mathrm{Re}(k_l)+i\,\mathrm{Im}(k_l)]$ of the period equation. The phase velocity ($c$) of the mode is equal to $\omega/\mathrm{Re}(k_l)$, while its attenuation ($Q^{-1}$) is equal to $2\,\mathrm{Im}(k_l)/\mathrm{Re}(k_l)$, then the group velocity is obtained by numerical differentiation from $c_g = \partial\omega/\partial\,\mathrm{Re}\{k_l\}$.

## 3. Numerical results

Dispersion, attenuation and excitation of the fundamental guided waves generated by monopole, dipole and quardrupole sources are analyzed. The multipole separation $r_0$ is 1 cm, while the borehole radius $a$ is 10 cm. The phase and group velocity of the modes are normalized with respect to the velocity of the bore fluid. In the follow results, we select the parameters as $\phi = 0.15$, $\kappa_0 = 5$ mD, $K_b = 16\times10^9$ N/m$^2$ $K_s = 38\times10^9$ N/m$^2$, $\mu_b = 14.61\times10^9$ N/m$^2$, $\alpha_\infty = 3.67$, $\eta = 0.001$ Pa.s, the density of grains was taken as $\rho_s = 2650$ kg/m$^3$. The density and Lamé constant of water were assumed to be $\rho_f = 10^3$ kg/m$^3$ and $K_f = 2.25\times10^9$ N/m$^2$. In all curves of excitation, z is equal to 3.0 meters.

A usual monopole (axisymmetric) source exites two types of waves. They are the Stoneley waves and the pseudo-Rayleigh modes (Fig.1, we only show the fundamental modes, the higher modes is not plotted). The Stoneley wave (Sto.) has no cutoff frequency. Its phase velocity starts at zero frequency with the so-called “tube wave” velocity and slightly reversed dispersion. The pseudo-Rayleigh mode (pR1) have a low cutoff frequency. Its phase and group velocity starts at the formation shear velocity and deceases with frequency toward the bore fluid velocity. The attenuation exhibits a strong intermediate maximum, correlated with the Airy phase (i.e., the minimum group velocity) of the mode and associated with the maximum of excitation of the mode. In fig.1a, we notice that both phase and group velocities of Stoneley wave and pseudo-Rayleigh mode are almost independent of the characteristic squirt flow length. However, an interesting aspect of Fig.1b is that the effect of the squirt flow is to increase the attenuation of Stoneley waves and pseudo-Rayleigh mode, relative to the no squirt flow case. Fig.1c shows that excitations of Stoneley waves and pseudo-Rayleigh mode are decreased due to the existence of squirt flow. Fig.2 displays the synthetic waveforms at three different receivers at increasing distance from the transmitter with a 8kHz(a) and 3kHz(b) source center frequency. $R$ is set to 2mm. At the low frequency, the fundamental mode-Stoneley wave is excited strongly than other waves, so the full waveform is mainly Stoneley wave. We can easy find that the maximum amplitudes of Stoneley waves (Fig.2b) and pseudo-Rayleigh mode (Fig.2a) are smaller than those of no squirt flow case. This property is consistent with the excitation curves in figure.1c. In Fig.2b, we also can find that arrival time of Stoneley waves is almost same for that based on BISQ model and Biot model. It is consistent with the dispersion curves in fig.1a.

Figure 3 shows the dispersion, attenuation and excitation curves of the fundamental mode (referred to as the flexural mode) excited by dipole point sources. The flexural waves (Flex.) are associated with a pure bending of the borehole. Its phase velocity ranges from the formation shear velocity toward that of a Scholte wave (i.e., a Stoneley wave at a fluid-solid interface) at high frequencies. The excitation curves have a peak associated with the Airy phase of the flexural mode. We also find that the dispersion curve of flexural mode is almost independent of the characteristic squirt flow length, however, the attenuation coefficient is increased due to the squirt flow, and the excitation is decreased due to the squirt flow. Fig.4 displays the synthetic waveforms with a 8kHz(a) source center frequency. $R$ is set to 2mm. In fig.4, we can find that arrival time of flexural waves is the same for that based BISQ model and Biot model. The dispersion curve in fig.2a also confirms it. The maximum amplitudes of flexural waves based on BISQ model are smaller than that of based on Biot model. It also clearly shows that shear waves ($S$) is not affected by the squirt flow.

Similar to the dipole point sources, the quadrupole point sources excite a fundamental mode (referred to as the screw mode). From figure 5, we can find the dispersion, attenuation and excitation of the screw mode (Scr.) are analogous to those of the flexural mode (Fig.3) with a shift toward to higher frequency. We find that the dispersion of screw mode is almost independent of the characteristic squirt flow length;

only very small difference can be seen around cut-off frequency at low frequencies. The attenuation coefficient is increased due to the squirt flow. Comparison of Fig.1 , Fig.3 and Fig.5 clearly shows similarity among the dispersion, attenuation and excitation of the pseudo-Rayleigh mode, flexural mode and screw mode. In fig.6, we can find that arrival time of screw waves is almost same for that based BISQ model and Biot model. The maximum amplitudes of screw waves based on BISQ model are smaller than that of based on Biot model. We can find that shear wave (*S*) is not affected by the squirt flow again.

## 4. Discussion

BISQ model predicts that attenuation of fundamental guided waves, which propagation in borehole embedded in porous formation, increases due to the squirt flow in saturated rock. The velocity dispersions of guided waves are not sensitive to the characteristic squirt flow length. The attenuation coefficients are sensitive to the characteristic squirt flow length, which encourages us to estimate the squirt flow length from the guided waves. It is of great interest in the inversion of guided waves (e.g. Stoneley wave, flexural wave and screw wave) attenuation from the acoustical logging data for the estimation of the characteristic squirt flow length. Of course, in the field, attenuation is much more difficult to measure than is velocity. Winkler et.al. [11] performed laboratory experiments to study borehole Stoneley waves, and their study shows the laboratory results are in excellent agreement with theoretical predictions based on Biot model. Their experimental porous rock sample (Berea S.S.) has intermediate value of the permeability ($220mD$). However, in throughout the present paper, we assume that the squirt flow phenomenon only occurred in those saturated porous rocks with low permeability (we may say $\kappa_0 < 100mD$).

## 5. Concluding remarks

We focus on investigation of the effects of the squirt flow on the fundamental guided waves propagation in borehole surrounded by saturated porous media with squirt flow. The numerical full waveforms show that the squirt flow influence on the amplitude of guided waves. Attenuations of Stoneley waves, pseudo-Rayleigh waves, flexural waves, and screw waves are enhanced due to the existence of squirt flow, excitations of those fundamental guided waves are decreased due to the existence of squirt flow, and velocity dispersions of guided waves are almost independent of squirt flow. It is possible to estimate the characteristic squirt flow length by attenuation coefficients of guided waves from acoustical logging data.

**Acknowledgements**

This work was supported by the National Natural Science Foundation of China under Grant No.40074032. The authors would like to thank professor Hu Hengshan of Harbin

Institute of Technology. And Dr. Tang Xiaoming of the Huston Technology Center is also appreciated.

**Figures captions**

Fig.1 Dispersion curves (a), attenuation coefficients (b) and excitation (c) of guided waves excited by monopole point source. Velocity is normalized by velocity of borehole fluid.

Fig.2 The full waveforms excited by monopole point source at axis of borehole.
(a) The source center frequency is 8kHz. (b) The source center frequency is 3kHz.
The characteristic squirt flow length is set to 2mm.

Fig.3 Dispersion curves (a), attenuation coefficients (b) and excitation (c) of guided waves excited by dipole point sources.

Fig.4 the full waveforms excited by dipole point sources. The source center frequency is 8kHz. The characteristic squirt flow length is set to 2 mm.

Fig.5 Dispersion curves (a), attenuation coefficients (b) and excitation (c) of screw waves excited by quadrupole point sources.

Fig.6 The full waveforms excited by quadrupole point sources. The source center frequency is 10kHz. The characteristic squirt flow length is set to 2 mm.

Fig.1

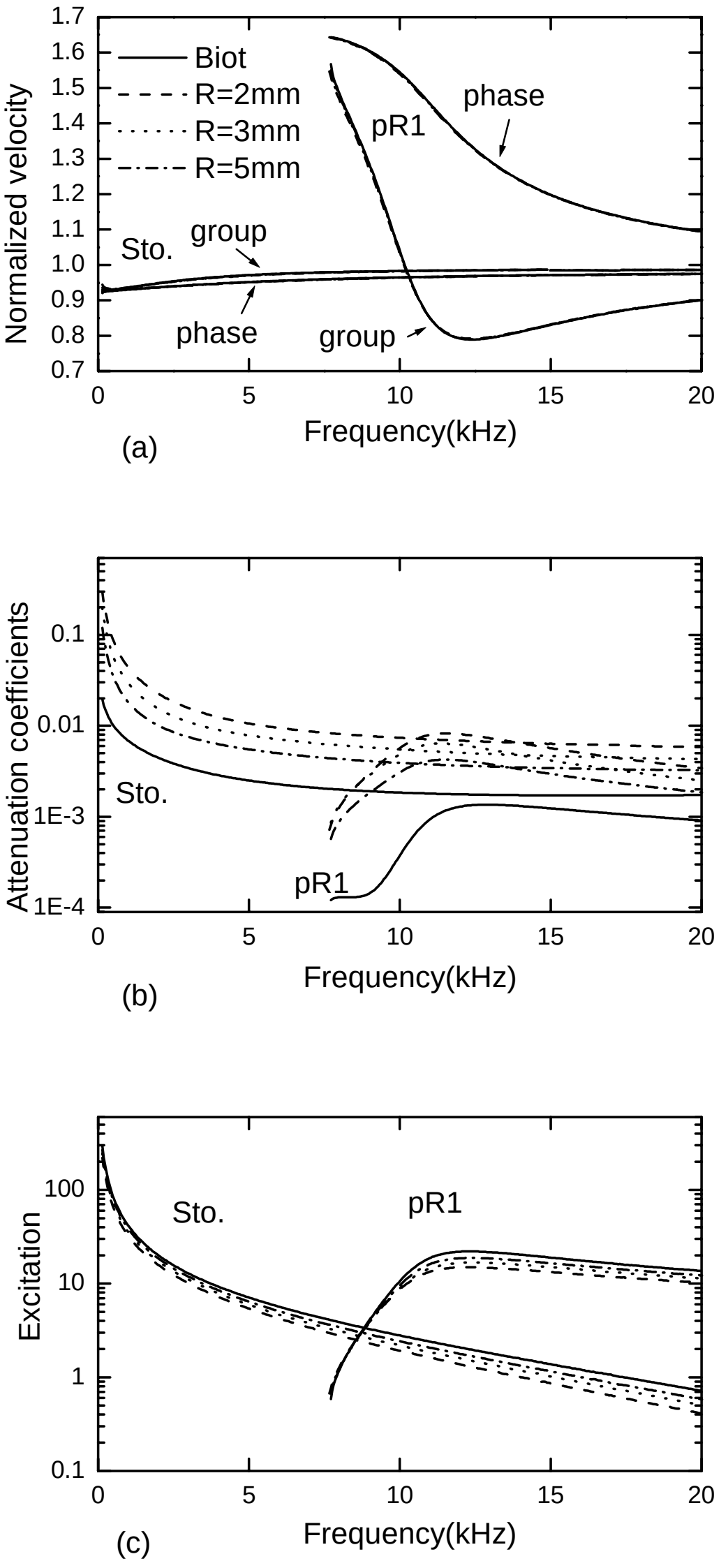

Fig.1 Dispersion curves (a), attenuation coefficients (b) and excitation (c) of guided waves excited by monopole point source. Velocity is normalized by velocity of borehole fluid.

Fig.2

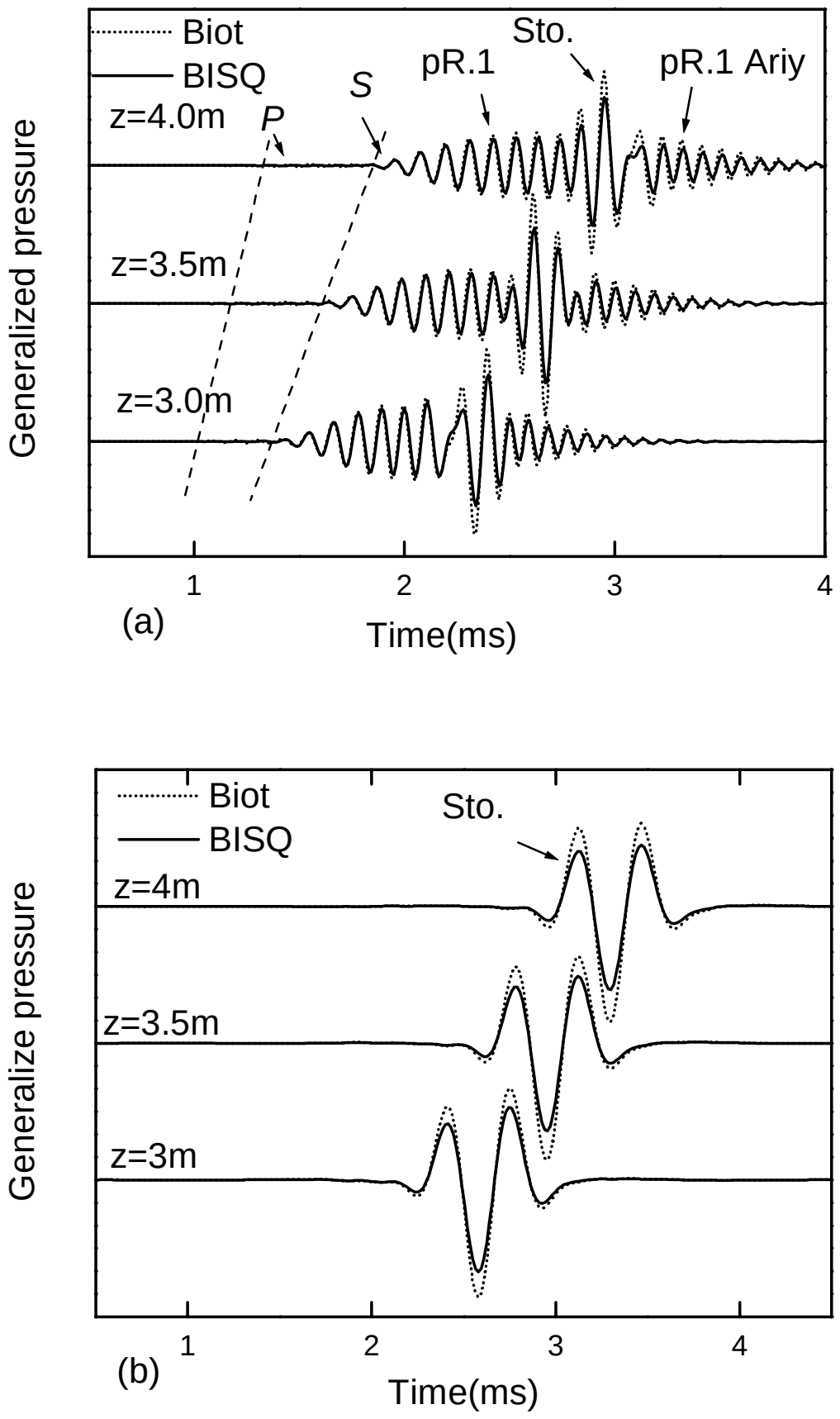

Fig.2 The full waveforms excited by monopole point source at axis of borehole. (a) The source center frequency is 8kHz. (b) The source center frequency is 3kHz. The characteristic squirt flow length is set to 2mm.

Fig.3

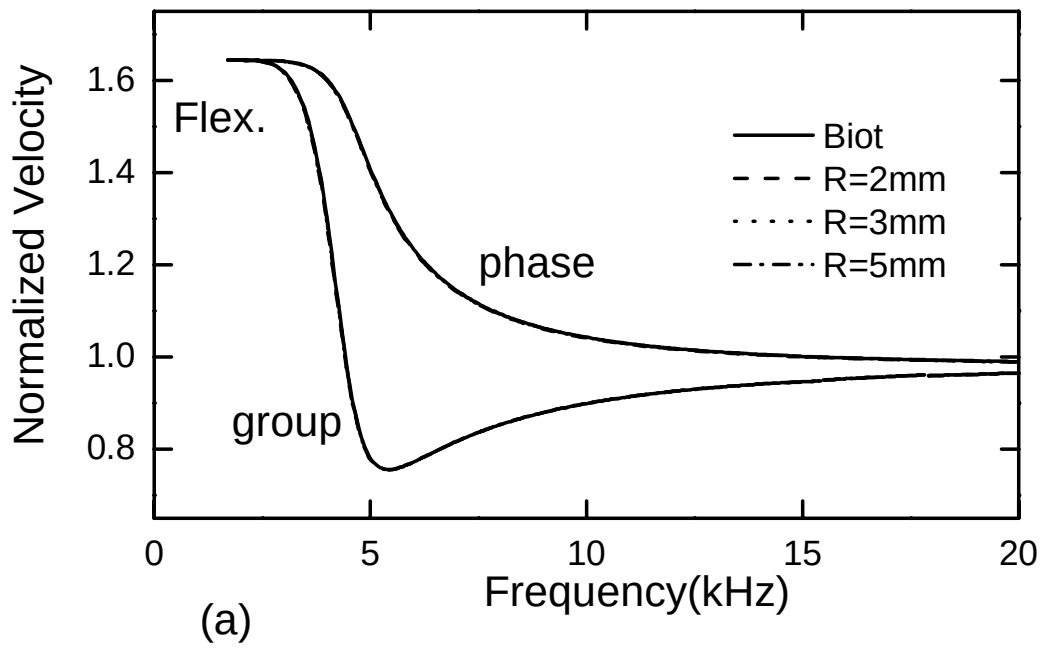

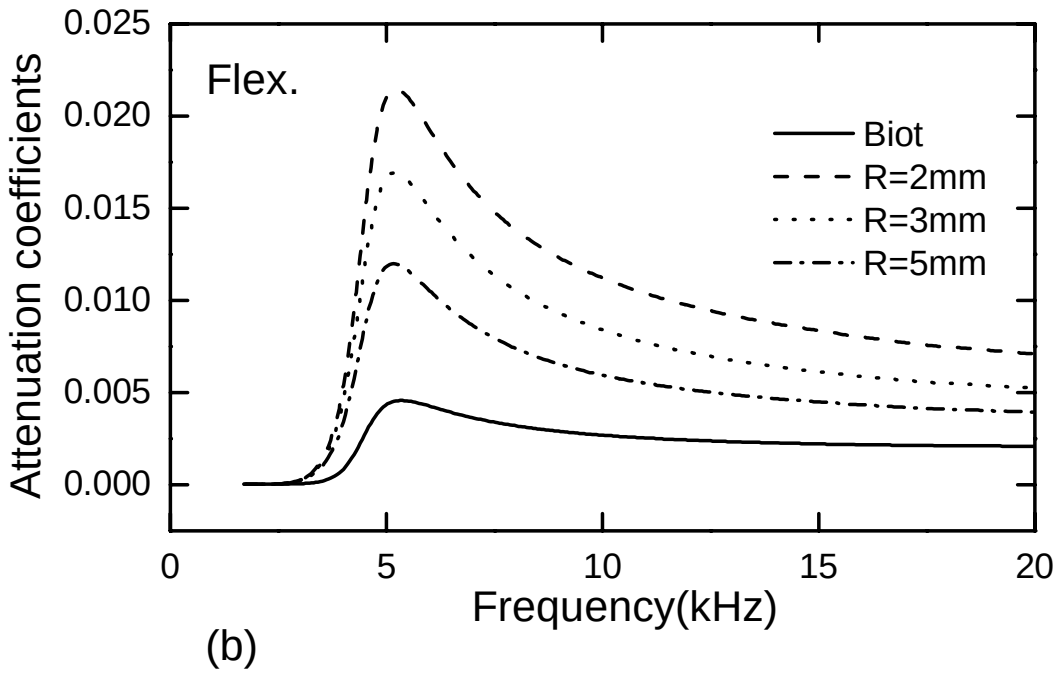

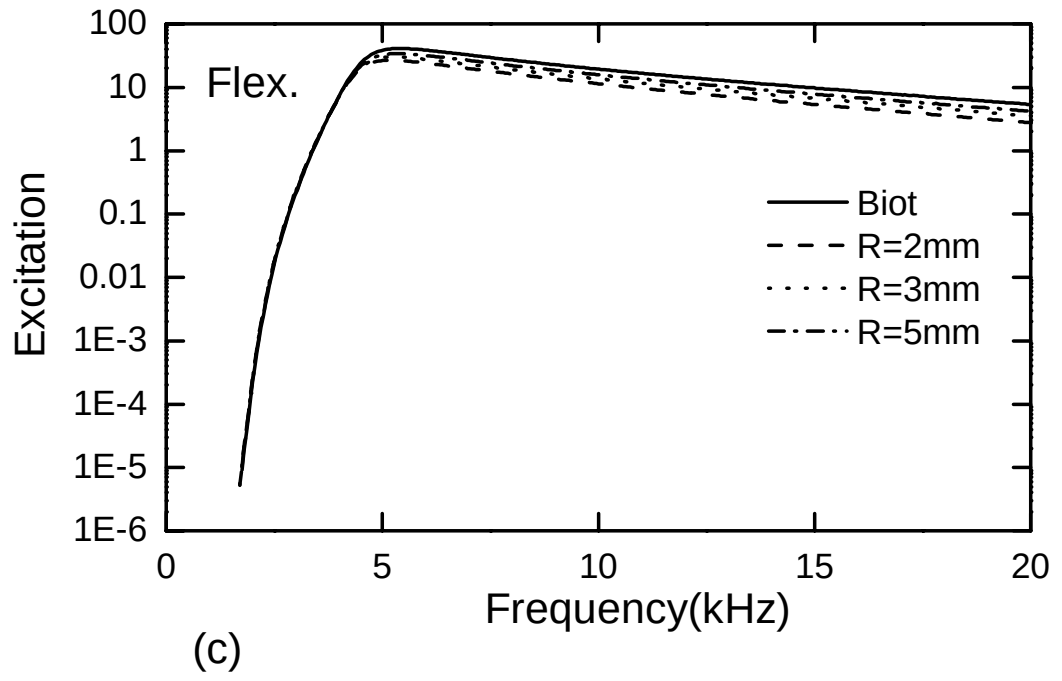

Fig.3 Dispersion curves (a), attenuation coefficients (b) and excitation (c) of guided waves excited by dipole point sources.

Fig.4

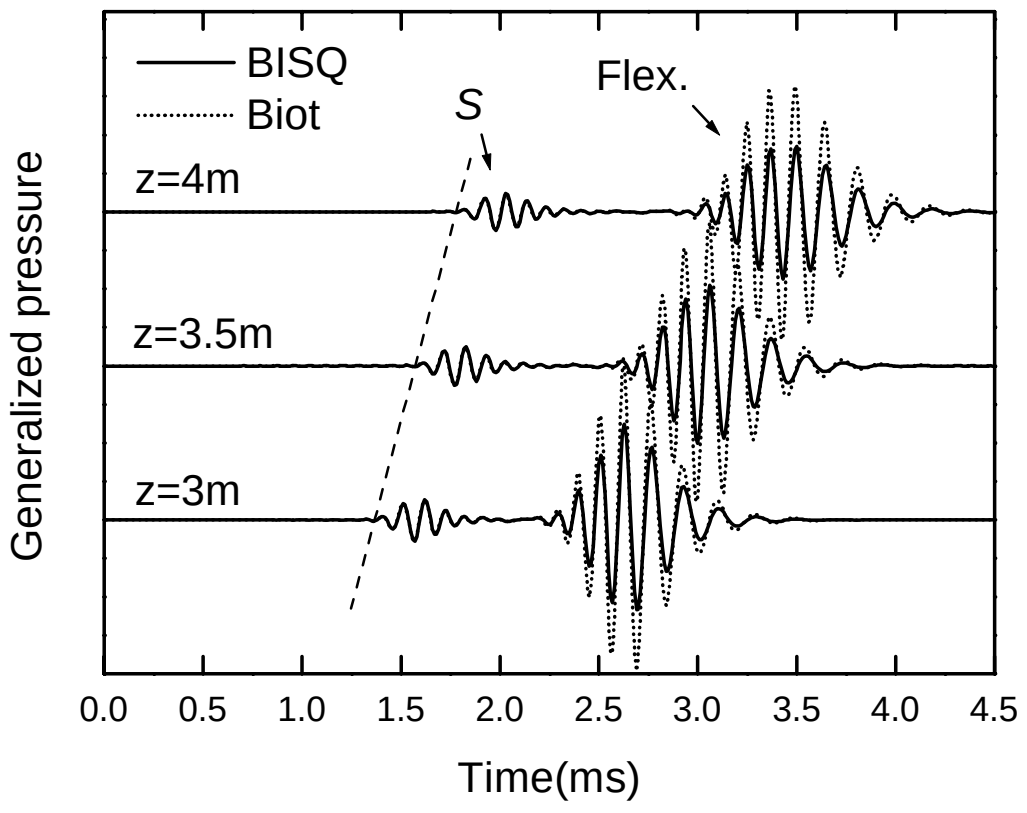

Fig.4 the full waveforms excited by dipole point sources. The source center frequency is 8kHz. The characteristic squirt flow length is set to 2 mm.

Fig.5

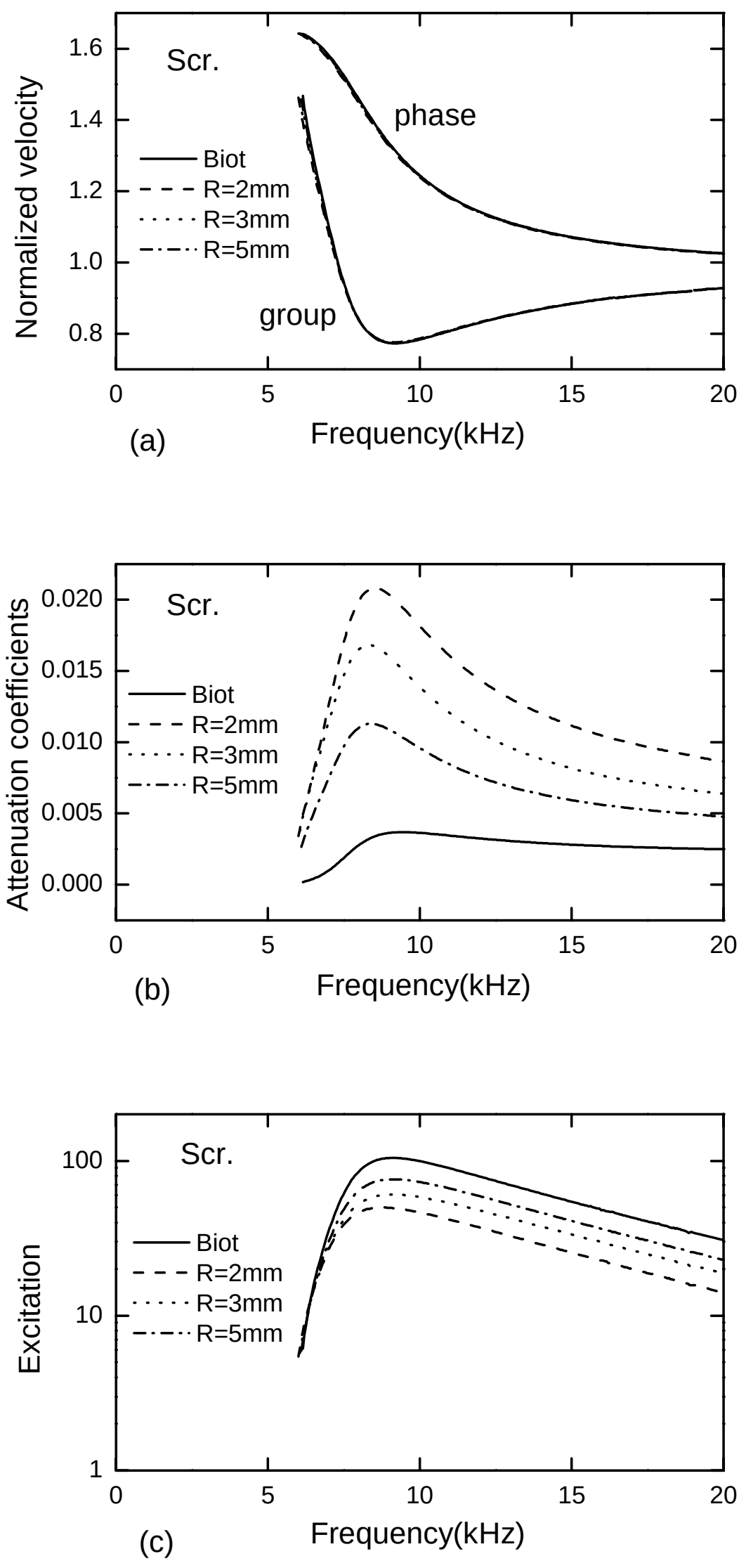

Fig.5 Dispersion curves (a), attenuation coefficients (b) and excitation (c) of screw waves excited by quadrupole point sources.

Fig.6

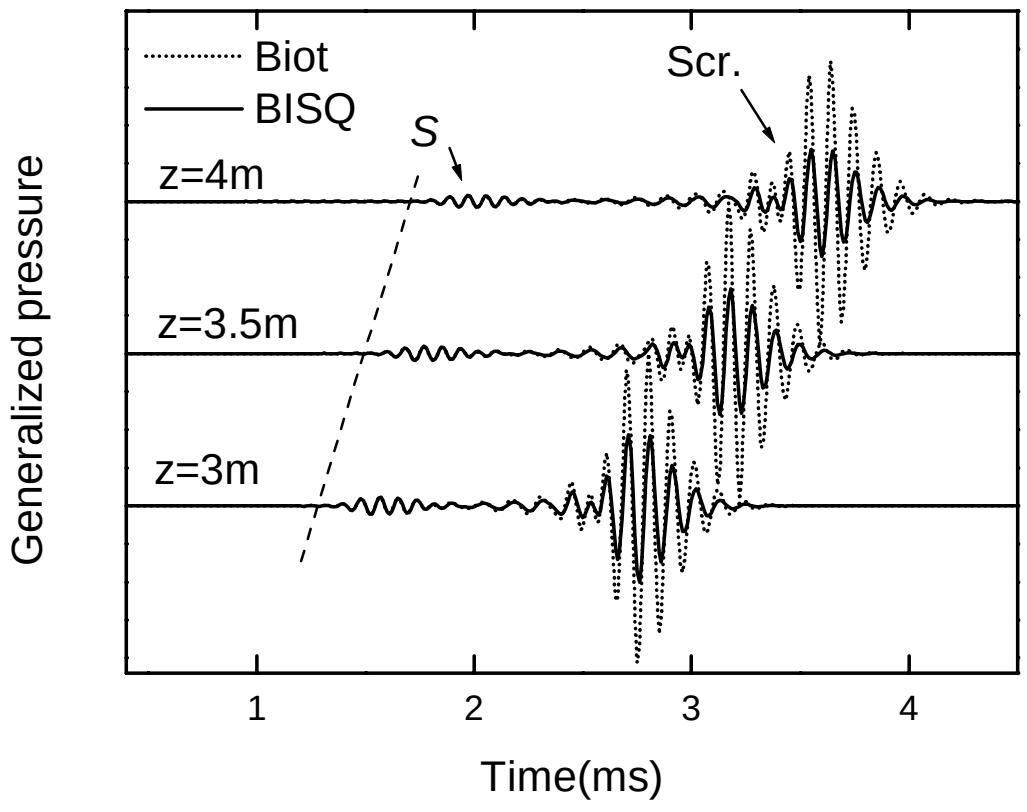

Fig.6 The full waveforms excited by quadrupole point sources. The source center frequency is 10kHz. The characteristic squirt flow length is set to 2 mm.